\documentclass[fleqn,usenatbib,useAMS]{mnras}

\usepackage{graphicx}	
\usepackage{amsmath}	
\usepackage{amssymb}	
\usepackage{multicol}        
\usepackage{bm}
\usepackage{pdflscape}	

\usepackage[T1]{fontenc}
\usepackage{ae,aecompl}

\usepackage{txfonts}

\usepackage{color}

\title[Assembly and structure of cluster haloes]{Assembly History and Internal Structure of Cluster Cold Dark Matter Haloes}

\author[Q. Chen et al.]{
Qingxiang Chen,$^{1,2,3}$\thanks{E-mail: chenqingxiangcn@gmail.com} 
Shihong Liao,$^{2}$
Jie Wang,$^{2,4,5}$ 
Liang Gao$^{2,4,5,6}$\\
$^{1}$International Centre for Radio Astronomy Research (ICRAR), The University of Western Australia, 35 Stirling Highway, Crawley, WA 6009, Australia\\
$^{2}$Key Laboratory for Computational Astrophysics, National Astronomical Observatories, Chinese Academy of Sciences, Beijing 100101, China\\
$^{3}$Yale-NUS College, Singapore, 138527, Singapore\\
$^{4}$School of Astronomy and Space Science, University of Chinese Academy of Sciences, Beijing 100049, China\\
$^{5}$Institute for Frontiers in Astronomy and Astrophysics, Beijing Normal University, Beijing 102206, China\\
$^{6}$School of Physics and Microelectronics, Zhengzhou University, Zhengzhou 450001, China}

\date{Last updated 2016 Nov 17}

\pubyear{2023}

\begin{document}
\label{firstpage}
\pagerange{\pageref{firstpage}--\pageref{lastpage}}
\maketitle

\defcitealias{Wang:2011a}{W11}


\begin{abstract}
We use the Phoenix simulations to study the mass assembly history and internal structures of cluster dark matter haloes ($M_{200} \gtrsim 5\times 10^{14} h^{-1}{\rm M}_\odot$). We confirm that cluster haloes grow inside-out, similar to galactic haloes. Major merger events dominate the growth of the internal region and minor mergers/diffuse accretion shape the outskirts. However, compared to galactic haloes, cluster haloes tend to have a younger and more actively evolving inner region. On average, the majority of mass ($\ga 80$\%) in the inner region ($R< 0.1 r_{200}$) of Phoenix haloes is accreted after $z = 3$, while for galactic haloes, most mass in the central region has already been accreted before $z=6$. The density profiles of cluster haloes are less stable than those of galactic haloes over different radii. The enclosed mass within $50$ or $150$ kpc of all Phoenix haloes evolves substantially in the past ${\sim} 7$ Gyr, while galactic haloes remained stable during the same period. We suggest that the relatively younger and more active state explains the various observations of cluster haloes, especially in central regions.
\end{abstract}


\section{Introduction}\label{sec:intro}

In the cosmological constant ($\Lambda$)--cold dark matter ($\Lambda$CDM) framework of galaxy formation \citep[e.g.][]{White:1978a,White:1991a}, dark matter first hierarchically forms virialized structures, the so-called dark matter haloes. Baryons are drawn into the gravitational potential wells formed by dark matter haloes. Once inside, they undergo shock-heating, followed by radiative cooling, leading to the formation of stars. This star formation is influenced by a sequence of feedback processes. Ultimately, these processes culminate in the creation of the luminous galaxies that we observe in the Universe today. Dark matter haloes serve as the backdrop for galaxy formation and evolution, exerting their influence through gravitational forces. Consequently, investigating the assembly and internal structures of dark matter haloes plays a pivotal role in comprehending the intricate processes underlying galaxy formation and evolution.

Among them, haloes of galaxy cluster with typical virial masses $M_{200} \ga 10^{14}$ $h^{-1}{\rm M}_\odot$ are the most massive gravitationally bound structures and they provide important information for tracing the matter distribution of the Universe \citep[see e.g.][for reviews]{Voit:2005a,Allen:2011a,Kravtsov:2012a}. As such, their mass fluctuations are sensitive to cosmological parameters, and serve as a good tool to constrain cosmological models. For example, the abundance of clusters can be used to constrain cosmological parameters such as $\Omega_{\rm m}$, $\sigma_8$, the equation of state for dark energy, and neutrino mass \citep[e.g.][]{Vikhlinin:2009a,Mantz:2015a,Planck:2016a,Bocquet:2019a}. Besides, as cluster haloes are the host for tens to hundreds of galaxies, their assembly history, structure and dynamics influence the evolution of galaxies in them. Especially, recent studies find that (proto)cluster environment may have contributed considerable cosmic star formation budget in the high redshift Universe \citep[e.g.][]{Casey:2016a,Chiang:2017a}. Understanding how cluster-scale structures have formed and evolved is thus important for both cosmology and galaxy formation theories. 

In observations, deriving physical properties of galaxy clusters often relies on the assumption that those systems are well relaxed. For example, as most of the mass is from dark matter components, mass estimation often relies on indirect methods except for gravitational lensing analysis, by measuring gas temperature, X-ray luminosity and gas mass and then scaling to total cluster mass \citep[e.g.][]{Pratt:2009a,Kravtsov:2006a}. These mass estimators are accurate only in virialization and hydrodynamic equilibrium conditions, so that assumptions underneath the scaling relations are met. However, clusters form much later than galaxies in the $\Lambda$CDM bottom-up assembly paradigm. Unlike galactic haloes, the dynamic state of clusters remains complex, and the relaxation requirement is not always guaranteed. Indeed, different methods of mass estimation (e.g. through hot gas measurement in X-rays; from analyzing distortions caused by cluster lenses; by observing the dynamics of surrounding galaxies etc.) can have significant offset, which may due to the complex dynamic state in clusters. Thus, how to identify such non-virialized clusters has been an important task in cluster mass reconstruction studies \citep[see e.g.][and references therein]{Rumbaugh:2018a,DeLuca:2021a}. Furthermore, the estimation of the density profiles of clusters is uncertain and depends on the assumption of the cluster's dynamic state \citep{Corless:2007a,Oguri:2010a,Sereno:2010a}. To have a better understanding of the above observational facts, we need to know clearly the assembly history of cluster halo and how it affects on the internal structure.

There is extensive literature to understand the assembly history of dark matter haloes using numerical simulations (see e.g. \citealt{Frenk:2012a} and \citealt{Zavala:2019a} for recent reviews). High-resolution resimulations of single objects bring us a detailed view of a halo's assembly history. Especially, with the Aquarius simulation \citep{Springel:2008a}, \cite{Wang:2011a} \citepalias[hereafter][]{Wang:2011a} studied in detail the assembly history and structure of Milky Way-sized dark matter haloes by tracing the history of all particles in a $z = 0$ halo to find out when and how they were accreted into galactic haloes. The resolution of Aquarius made it possible to track the whole assembly history of each particle, and hence gave the most detailed investigation of the structure formation and merger histories. They show a clear `onion-like' inside-out growth picture for galactic haloes, and find that the galactic haloes accrete most of the mass from the not-so-active mergers. In general, accretions happen at early times, resulting in a very stable structure at redshift 0. \citet{Genel:2010a} did a similar study with the Millennium simulations and found that smooth accretion is important for the growth of all haloes covering the ranges of $10^9 -10^{15}{\rm M}_\odot$. 
However, a more detailed study of cluster haloes based on high resolution resimulation, like the analysis of galactic haloes in \citetalias{Wang:2011a}, is still needed. Such studies will offer us a deeper understanding in the virialization and dynamic states of cluster-sized haloes.

In this work, we investigate whether similar conclusions of \citetalias{Wang:2011a} hold in high-resolution cluster dark matter haloes using the Phoenix simulations \citep{Gao:2012a}. These are high-resolution re-simulations of nine cluster haloes originally selected from the Millennium simulation \citep{Springel:2005a}. The Phoenix simulations have comparable number of high-resolution particles as the Aquarius simulations and are performed with similar numerical codes, and thus they are particularly suitable for us to compare the results with those of \citetalias{Wang:2011a} directly. By closely tracing the behaviour of mass particles, we hope to better understand cluster haloes' complex mass assembly history and inner structures.

This paper is organized as follows. In Section \ref{sec:simulation}, we introduce the details of simulations used in this work. In Section \ref{sec:history}, we study the assembly history and accretion modes of cluster haloes. In Section \ref{sec:structure}, we investigate the evolution of inner structures and density profiles. We discuss the implications for observations in Section \ref{sec:discussion}. Section \ref{sec:summary} summarizes our main conclusions.

\section{Simulations}\label{sec:simulation}
In this work, we use the Phoenix simulations \citep{Gao:2012a} to study the assembly and structure of cluster haloes, and compare with those of galactic haloes from the Aquarius simulation suite \citep{Springel:2008a}. These two dark matter-only simulations both use the WMAP 1-year cosmological parameters: mass density $\Omega_{\rm m}=0.25$, cosmological constant $\Omega_\Lambda=0.75$, power spectrum normalization $\sigma_8=0.9$, primordial spectral slope $n_{\rm s}=1$, and Hubble constant $H_0=73\  {\rm km}\ {\rm s}^{-1}{\rm Mpc}^{-1}$ \citep{Spergel:2003a}. Both simulation sets were conducted using the \textsc{gadget-3} code, initially developed for the Aquarius Project. \textsc{gadget-3} represents an updated iteration of the publicly accessible \textsc{gadget-2} code \citep{Springel:2005b}. 
Haloes and subhaloes in simulations are identified using the friends-of-friends \citep[FOF,][]{Davis:1985a} and \textsc{subfind} algorithm \citep{Springel:2001a}, respectively. The merger trees for the final haloes are constructed by linking the FOF progenitors at each snapshot, and the merger trees for subhaloes are constructed as described in \citet{Springel:2005a}.

Throughout this paper, we consistently refer to $M_{200}$ as the virial mass of a halo, defined as the mass contained within a sphere whose average density is 200 times the critical value of the Universe. Concurrently, the associated virial radius is denoted as $r_{200}$. The coordinate of the deepest potential (i.e. the position of the most-bounded particle) is defined as the halo centre.

\subsection{Phoenix simulations}

In the Phoenix project, a total of nine haloes were selected from the Millennium simulation \citep{Springel:2005a}. These haloes encompass a range of virial masses spanning from $5\times 10^{14}$ $h^{-1}{\rm M}_\odot$ to $2\times 10^{15}$ $h^{-1}{\rm M}_\odot$. These haloes were individually re-simulated using a zoom-in technique, and are denoted as Ph-X, where X varies from A to I. To investigate numerical convergence, the Ph-A halo was subjected to simulations at four different mass resolution levels, labeled as level-1 (highest resolution) through level-4 (lowest resolution). For the remaining haloes, Ph-B through Ph-I, two resolution levels were performed, namely level-2 and level-4. It is worth noting that in this study, we primarily focus on the results from level-2 and level-4 simulations.

The level-2 simulations feature a mass resolution ($m_{\rm p}$) of approximately $10^6$ $h^{-1}{\rm M}_\odot$ and employ a softening length of $\epsilon = 0.32$ $h^{-1}{\rm kpc}$. In contrast, the level-4 simulations have a coarser mass resolution of around $10^8$ $h^{-1}{\rm M}_\odot$ and a larger softening length of $\epsilon = 2.8$ $h^{-1}{\rm kpc}$. Within the virial radius of each level-2 and level-4 halo, there are approximately $10^8$ and $10^6$ particles, respectively.

Each Phoenix simulation consists of a total of 72 snapshots, labelled from 0 to 71, with snapshot 71 corresponding to the redshift of 0. For more information on the Phoenix project including the basic halo properties and convergence studies, readers are referred to \cite{Gao:2012a}. 

\subsection{Aquarius simulations}

The Aquarius Project involved the re-simulation of six Milky Way-sized haloes, denoted Aq-A to Aq-F, in isolated environments within the lower resolution version of the Millennium II simulation \citep{Boylan-Kolchin:2009a}. These haloes exhibit virial masses ranging from $1\sim 2 \times 10^{12}$ $h^{-1}{\rm M}_\odot$. The Aq-A halo, in particular, was simulated at five different levels of mass resolution, while the remaining five Aquarius haloes were subjected to simulations at two lower resolution levels. In our study, we primarily focus on the results from the level-2 Aquarius simulations.

In the level-2 Aquarius simulations, which are central to our investigation, the mass resolution ($m_{\rm p}$) is approximately $10^4$ ${\rm M}_\odot$, the softening length is set at $\epsilon = 65.8$ ${\rm pc}$, and the number of particles contained within the virial radius is around $N_{200} \sim 10^8$. For more detailed information regarding the Aquarius project, readers are encouraged to refer to \cite{Springel:2008a}.

\section{Mass accretion history}\label{sec:history}

In this section, we delve into the mass growth history of the Phoenix haloes and draw comparisons with the Aquarius haloes. We focus on simulations conducted at level-2 resolution, which represents the highest resolution level common to all Phoenix and Aquarius haloes. It is worth noting that a comprehensive convergence analysis of Aquarius results has been previously conducted in \citetalias{Wang:2011a}. Given that both the Phoenix and Aquarius simulations share similar characteristics, including a comparable number of high-resolution particles and the use of the same numerical code, we anticipate that the Phoenix simulations should exhibit similar convergence behaviours.

\subsection{Mass growth}

\begin{figure}
    \centering
    \includegraphics[width=0.9\columnwidth]{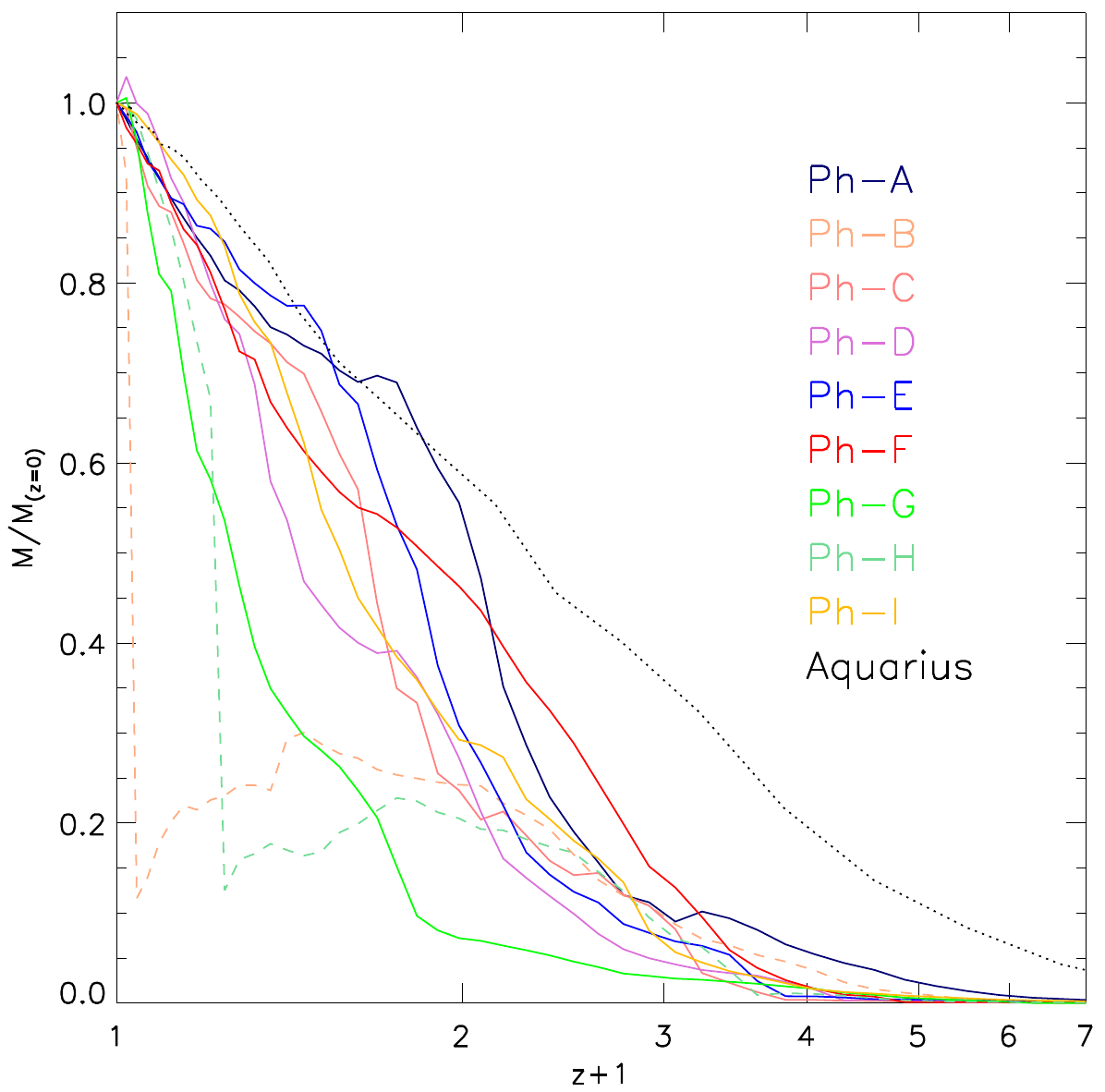}
    \caption{
	Mass growth history of Phoenix haloes. The horizontal axis is redshift and the vertical axis is the halo mass read from the merger tree's main branch, relative to the halo mass at $z=0$. Ph-B and Ph-H, which experience dramatic mass decrease and increase at $z<0.5$, are plotted as dashed lines, while the others are in the solid lines. The dotted curve is the averaged mass growth history for Aquarius haloes.
	}
    \label{fig:growth_ph}
\end{figure}

We first look at the mass growth history, $M_{200}(z)$, of nine Phoenix haloes by tracing the main branches of their merger trees. In Figure~\ref{fig:growth_ph}, we show their halo masses at different redshift in coloured lines. The horizontal axis is redshift, and the vertical axis is the main branch progenitor halo mass scaled by the mass at $z=0$. For comparison, we have incorporated the overall mass evolution of the Aquarius haloes into our analysis, as shown as the black dotted line. To achieve this, we computed the average mass across all seven haloes at various redshifts. Readers seeking specific details regarding the individual halo's mass evolution are encouraged to refer to Figure~1 in \citetalias{Wang:2011a}. 

The growth history for haloes indicated as solid lines, are in general similar, with rapid mass growth after $z\sim 1$. This is especially true for Ph-G. It has only $<10\%$ of its final mass at $z=1$, and has not accreted half of its final mass until $z\sim 0.18$. 

The two haloes, Ph-B and Ph-H, exhibit significant discontinuities in their mass growth histories, occurring notably at snapshot 69 ($z = 0.04$) and snapshot 62 ($z = 0.24$), respectively. Moreover, both of these haloes experience substantial mass losses as they evolve into redshifts below 1. They are marked with two dashed lines in Figure \ref{fig:growth_ph}. We have conducted an investigation to understand the origins of these conspicuous mass jumps, which are associated with mergers between the primary halo and a smaller, yet denser halo. In such instances, the merger tree algorithm faces challenges in correctly identifying the main progenitor. This is primarily due to the fact that the most bound particle does not necessarily belong to the most massive halo.\footnote{This problem arises because the distinction between haloes and subhaloes can become difficult in major merger scenarios, making methods that use temporal information crucial in such cases (as pointed out by \citealt{Behroozi:2013a, Behroozi:2015a} and references therein). Methods such as \textsc{hbt} \citep{Han:2012a} can be employed to improve this, but it is beyond the scope of this work.}

Interestingly, the Aquarius haloes do not encounter similar issues, possibly owing to their relatively limited involvement in major mergers that would significantly perturb their central regions in recent times. While it is conceivable to adopt an alternative definition and modify the merger tree algorithm to address this problem, such endeavors fall beyond the scope of this study.

Hence, for the entirety of this work, we have chosen to exclude these two haloes from our analysis. Nevertheless, it remains an intriguing avenue for future research to delve into the merging events that have impacted these particular haloes, shedding further light on their unique evolutionary trajectories.

It can be readily found from Figure \ref{fig:growth_ph} that the mass growth of clusters happens very late relative to galaxies. In general, more than 90\% of the final halo mass is accreted after $z=2$. Compared to the Aquarius haloes \citepalias[i.e. Figure 1 of][]{Wang:2011a}, the slopes of $M_{200}(z)$ for the Phoenix haloes are in general steeper. This is expected in the hierarchical formation model, which predicts more recent growth for cluster haloes compared to galactic ones \citep[see e.g.][]{NFW:1996,Mo:2010}. Late assembly suggests that cluster haloes may not be fully virialized compared to galactic haloes. We investigate the mergers and accretions in more detail in the next sections.

\subsection{Merger ratio profile}

\begin{figure*}
    \includegraphics[width=0.9\textwidth]{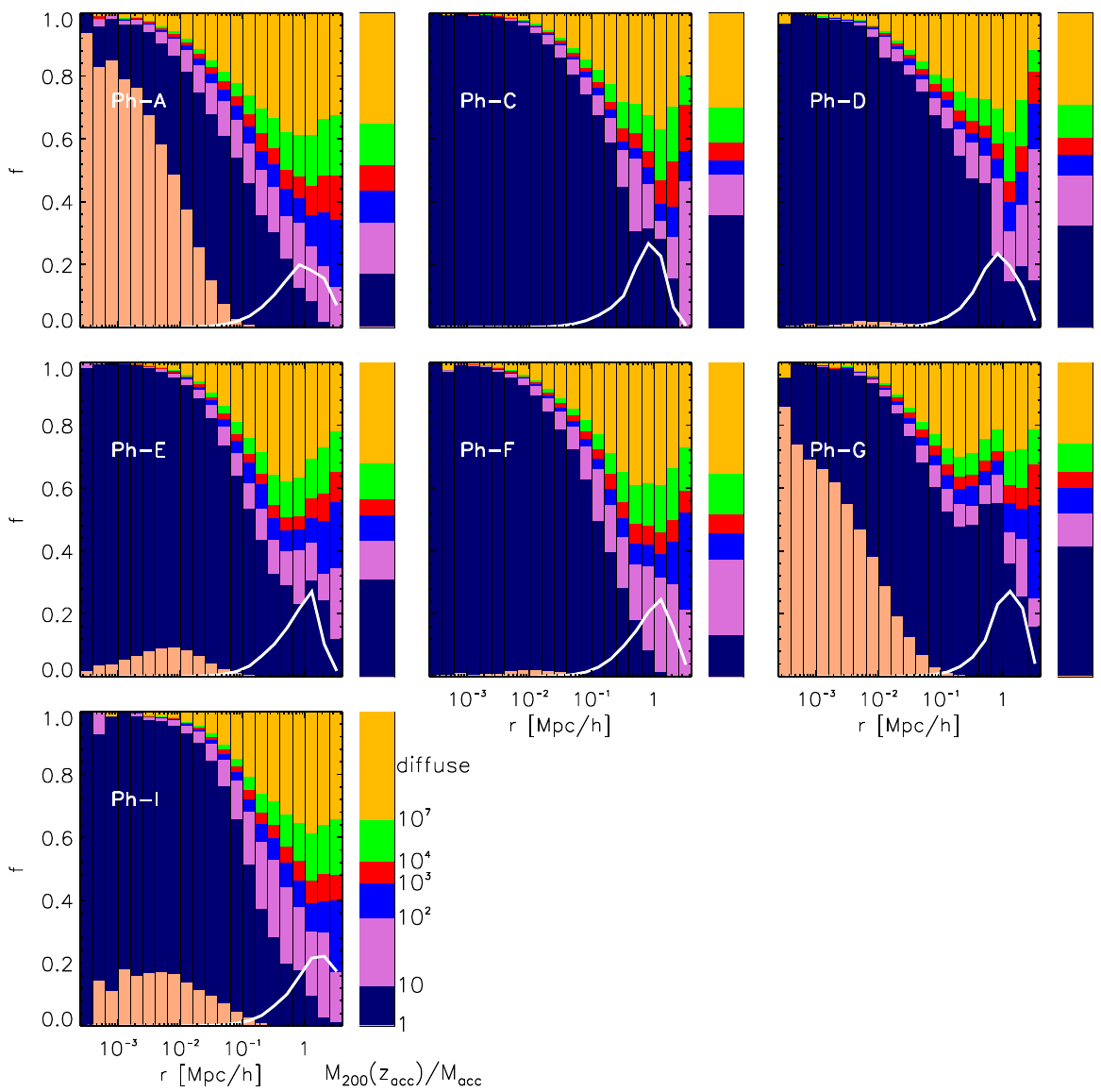}
    \caption{
	The radial dependence of Phoenix halo merger ratios. Colours are coded by the division between a particle's two parent halo masses when the (first) accretion event happens for this certain particle. The horizontal axis is the spherical radius. The vertical axis records the fraction of particles with different merger ratios at a given radius bin. The colour bars on the right of each subplot show the fraction of total mass contribution fraction of each merger ratio for the halo. The white curve in each plot gives the fraction to the total plotted mass in each radius bin. For the convenience of comparison with \citetalias{Wang:2011a}, we stop tracing merger events at $z = 6$ in this analysis. The diffuse particles are those do not belong to any resolved structures when they (first) fall on the main branch. Very few particles of $z>6$ (colour coded in beige, but not labeled in the colour bar for better clarity) contribute to the final total mass.}
    \label{fig:merger_ratio_ph}
\end{figure*}

\begin{figure*}
    \includegraphics[width=0.85\textwidth]{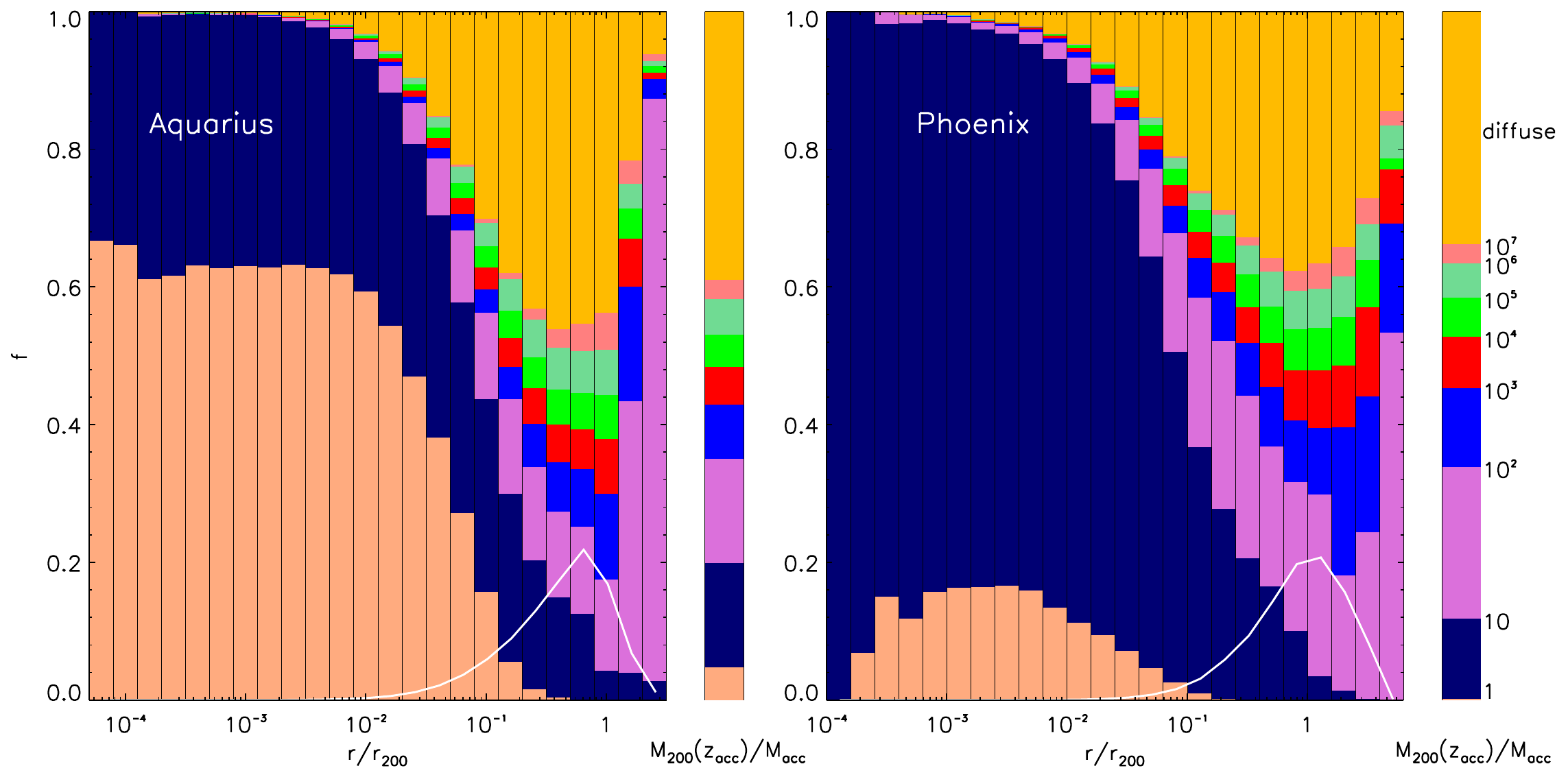}
    \caption{
	The stacking merger mass ratio profiles of Aquarius haloes (left) and Phoenix haloes (right). As in Figure~\ref{fig:merger_ratio_ph}, we stop tracing merger events to earlier than $z=6$. The white curves show the total mass fraction of the halo mass in each shell. Every particle's distance is normalized to $r_{200}$ of its host halo before being stacked. For both Aquarius and Phoenix haloes, the bigger merger ratio is more likely to happen towards outskirts. The main difference between the two is the $z>6$ bins (not labeled in the colour bar for better clarity), as Aquarius haloes accrete relatively more mass at high redshifts compared to Phoenix haloes. Other than this, haloes at the two mass scales show strong self-similar patterns. Major merger events with merger ratio $1\sim 10$ contribute to the central-most regions. In outskirts, smaller merger events take place more often. The diffuse accretions also contribute mostly to outer regions.}
    \label{fig:merger_ratio_Aq2Ph2}
\end{figure*}

\begin{figure}
    \centering
    \includegraphics[width=0.9\columnwidth]{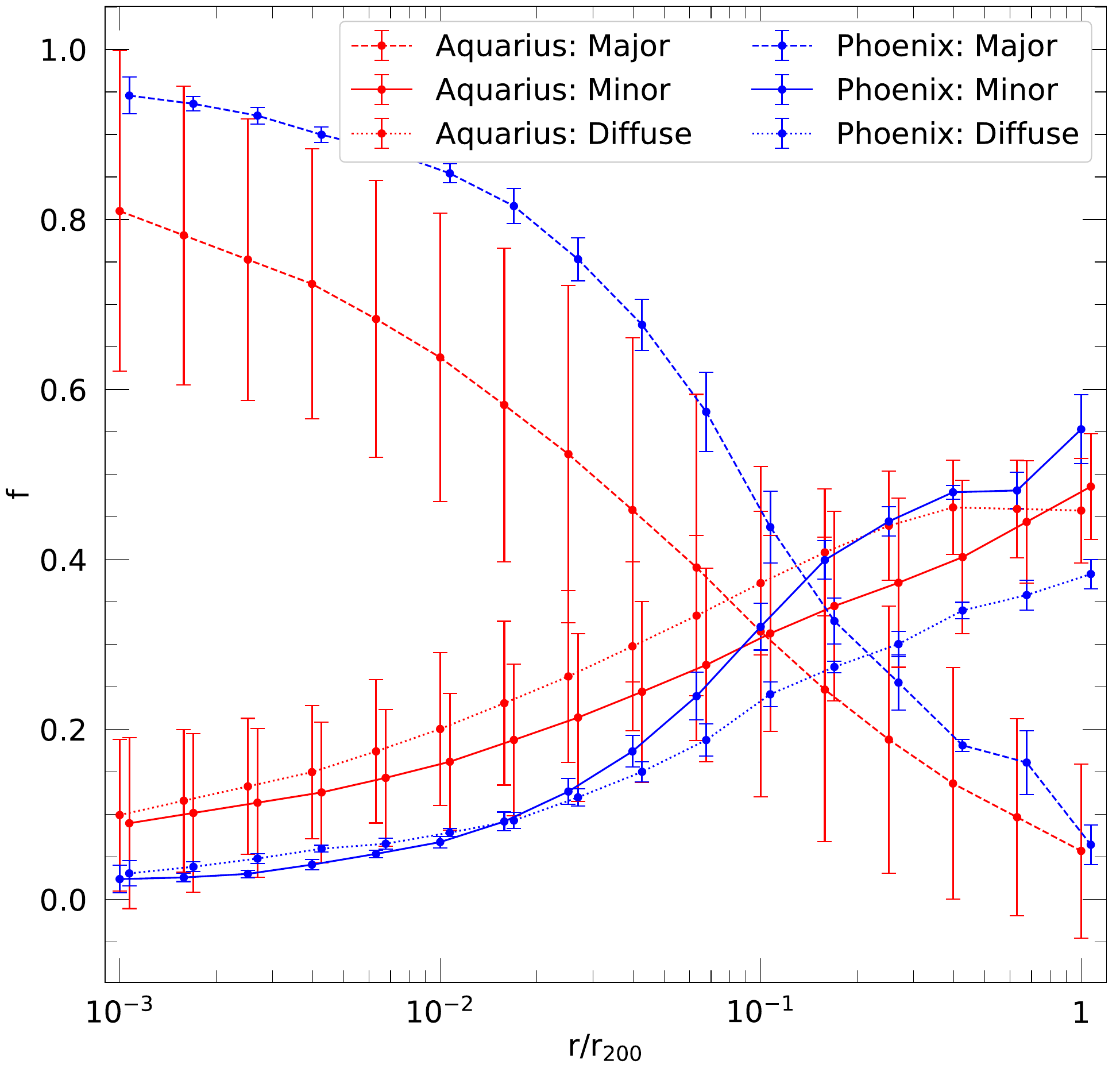}
    \caption{
	The radial profiles of accretion modes for Aquarius (red) and Phoenix (blue) haloes. The major mergers, minor mergers, and diffuse accretion are marked with dashed, solid, and dotted lines. Error bars show the standard deviations among haloes. Lines for Phoenix major merger, Aquarius minor merger, Phoenix diffuse are right shifted slightly for better illustration. We see that major merger dominates the inner regions, while minor merger and diffuse situation dominate the outskirts, for both Aquarius and Phoenix haloes. However, the trends are more obviously seen for Phoenix haloes. Minor mergers are less dominant than diffuse accretions for Aquarius haloes, while for Phoenix haloes they are more dominant.
	}
    \label{fig:merger_path_profile_Aq2Ph2}
\end{figure}

In accordance with the methodology outlined by \citetalias{Wang:2011a}, we investigate merger event patterns for particles residing at varying radii within each Phoenix halo. At the redshift $z=0$, we meticulously investigate the history of each particle located in distinct radial shells. This investigation is conducted by tracking the particle's history across different snapshots based on its unique particle ID. We discern the pivotal moment when each particle was first accreted into the main halo progenitor of the Phoenix halo, which serves as the accretion time for this particle.\footnote{Note that there are different definitions for particle accretion time (see \citetalias{Wang:2011a} for details) and different halo definitions (e.g. FOF versus spherical overdensity), which can quantitatively affect the results presented in this work. FOF haloes may on occasion suffer from the `FOF bridge' effect which misconnects two close haloes. To provide a straightforward comparison between Phoenix and Aquarius haloes, we adopt the same definitions of haloes and particle accretion events as \citetalias{Wang:2011a}.} Furthermore, as we identify the point of accretion into the main halo progenitor, we also record the mass of the halo ($M_{\rm acc}$) responsible for bringing the particle into the main halo progenitor. This enables us to calculate the mass ratio between the main halo progenitor and the contributing halo, denoted as ${\rm ratio} = M_{200}(z_{\rm acc})/M_{\rm acc}$. Under this definition, a ratio closely approximating unity signifies that the particle was introduced into the halo progenitor through a major merger event, whereas a ratio significantly exceeding unity indicates that the particle entered through a minor merger event.

In Figure \ref{fig:merger_ratio_ph}, we examine the radial distribution of merger mass ratios across seven Phoenix haloes. Our approach involves categorizing the mass ratios into distinct bins, each represented by a unique colour on the colourbar within each panel. Subsequently, we plot the fractions of particles at $z=0$ associated with various merger mass ratios within different radial bins. It is worth noting that if a particle lacks association with any resolved structure at the time of its accretion, we categorize this event as `diffuse'. A clump of mass must contain at least 20 particles to be resolved as a subhalo identified by \textsc{subfind}. Therefore, the category `diffuse' in principle includes those unresolved mergers. However, we use the term `diffuse' to refer to `unresolved mergers' and `smoothed accretion' without distinguishing between their physical contexts in this work. To maintain consistency with the analysis conducted by \citetalias{Wang:2011a}, we halt our trace-back analysis at $z=6$, designating particles already situated within the main halo progenitor at $z = 6$ as `$z>6$'. As indicated in Figure \ref{fig:growth_ph}, it is evident that only a small number of particles have already assumed their positions before the redshift of $z=6$.

The Phoenix haloes exhibit similar patterns to the Aquarius haloes, as depicted in Figure 3 of \citetalias{Wang:2011a}. Specifically, we observe that particles accreted prior to approximately $z\sim 6$ and those introduced through relatively major mergers, characterized by mass ratios falling in the range of $1\sim 10$, constitute the primary contributors to the central regions of the haloes. In contrast, particles from diffuse accretion are infrequently found in the inner regions but assume a more substantial presence in the outskirts. Minor mergers, characterized by mass ratios exceeding 10, also play a significant role in shaping the outer regions of the haloes. In general, mass accreted before the redshift of $z\sim 6$ does not make a substantial contribution.

A notable overarching trend in the merger ratio distribution is the correlation between larger merger ratios and greater radial distances from the halo centre. While there is considerable variability among haloes regarding particles already in place before $z\sim 6$, differences between the haloes are relatively minor. Notably, if we exclude particles that were already present before redshift 6, the patterns of merger ratios across all radii exhibit a high degree of uniformity across all seven Phoenix haloes. This echoes the universal unevolved subhalo mass function and merger ratio distributions found over a wide range of halo masses \citep[see e.g.][]{van_den_Bosch:2005,Han:2016,Han:2018,Dong:2022}.

We subsequently normalize the shell radii by the virial radius, denoted as $r_{200}$, for each individual halo. This normalization facilitates a meaningful comparison of the averaged radial fractions of merger ratio profiles between the Phoenix haloes and the Aquarius haloes, as showcased in Figure \ref{fig:merger_ratio_Aq2Ph2}. To ensure a fair assessment, the haloes within each project are averaged within each radius bin. Given that galactic haloes tend to form at earlier cosmic epochs, it is not surprising that the particles originating at redshifts exceeding 6 constitute a more significant fraction in comparison to cluster haloes. However, when we focus our analysis on the contributions from particles at lower redshifts and disregard the high-redshift particles, an intriguing revelation emerges: these two types of haloes actually exhibit strikingly similar merger ratio profiles. Specifically, the central regions predominantly consist of major mergers with ratios ranging from 1 to 10, while progressively larger merger ratios become more prevalent in the outskirts. Furthermore, diffuse particles primarily participate in mergers in the outskirts, where they exert a considerable influence.

To gain a deeper understanding of how different merger patterns vary with radius, we classify particles into three distinct accretion modes based on their merger mass ratios. Specifically, we define major mergers as events with mass ratios falling within the range of 1 to 10, minor mergers as events with mass ratios exceeding 10, and diffuse accretions as particles that were not associated with any resolved structures at the time of their accretion ($z_{\rm acc}$). In Figure \ref{fig:merger_path_profile_Aq2Ph2}, we present the mean fractions of these three accretion modes as a function of re-scaled radius ($r/r_{200}$) for both Aquarius and Phoenix haloes. It is important to note that in this analysis, we do not impose a redshift cutoff when tracing the particle histories. Consequently, particles accreted at redshifts exceeding 6 are also considered, and their merger ratios are calculated.

The comparison reveals similar merger patterns between the Aquarius and Phoenix haloes. Major mergers predominantly contribute to the centralmost regions, whereas minor mergers and diffuse accretions are more prevalent in the outskirts. Notably, in the case of Phoenix haloes, this trend is even more pronounced, with major mergers playing a larger role in the central regions compared to Aquarius haloes. Additionally, in the outskirts of Phoenix haloes, minor mergers contribute more than their Aquarius counterparts. Within the virial radius, diffuse accretion dominates minor mergers in Aquarius haloes, whereas for Phoenix haloes, the mass contribution from minor mergers exceeds that of diffuse accretion at radial distances beyond approximately $0.02 r_{200}$.

The findings presented above highlight that, on the whole, Phoenix haloes exhibit qualitatively similar patterns of structure formation as Aquarius haloes. However, there are notable quantitative distinctions in their merger pathways. In particular, it appears that merger events exert more pronounced effects on Phoenix haloes compared to their Aquarius counterparts. This effect is especially pronounced for particles located in the central regions, where the Phoenix haloes tend to have undergone more recent activity and experienced a higher frequency of major mergers.

\subsection{Accretion time profile}

\begin{figure*}
    \includegraphics[width=0.9\textwidth]{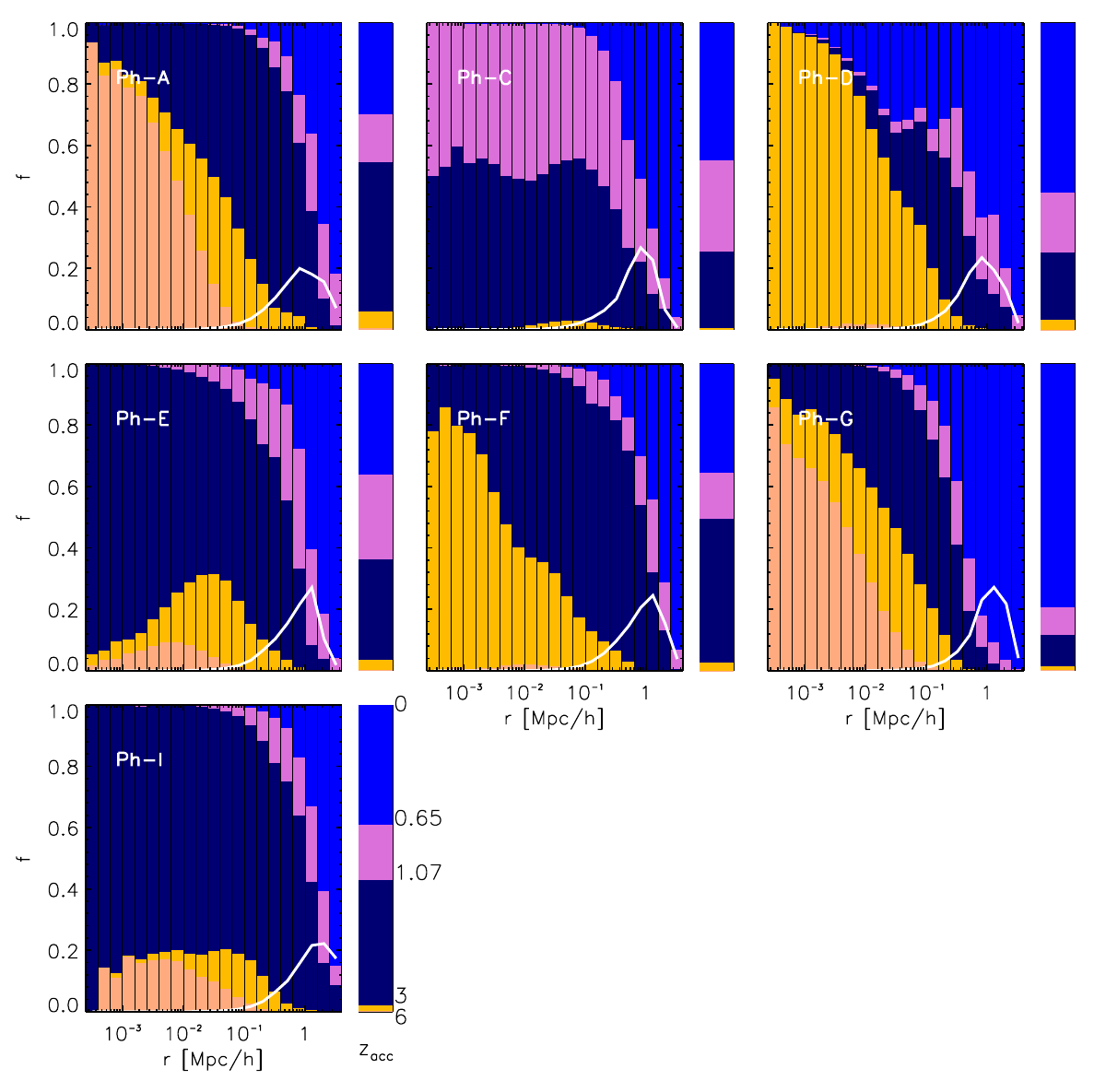}
    \caption{
	The radial profiles of Phoenix haloes' accretion redshifts. Like in Figure \ref{fig:merger_ratio_ph}, we split the halo in radial shells, and colour code the accretion redshift of particles in them. For each halo, the horizontal axis is the radius and vertical axis is the fraction of particles contributed in each redshift bin. The colour bar on the right to each subplot indicates the colour coding of redshift bins. For better clarity in the colour bar, particles of $z>6$ (in beige) are not labeled.
	}
    \label{fig:acc_time_profile_ph2}
\end{figure*}

Next, we turn our attention to examining the radial profiles of particle accretion times within Phoenix haloes. Analogous to the particle mass ratio profile discussed in the previous subsection, we segment the particle accretion redshift, denoted as $z_{\rm acc}$, into distinct bins. For each radius within each Phoenix halo, we calculate the proportion of particles falling into various $z_{\rm acc}$ bins. The resultant profiles for all seven Phoenix haloes are visually depicted in Figure \ref{fig:acc_time_profile_ph2}.

In a broad sense, all haloes demonstrate an inside-out formation pattern, characterized by early accretion of mass in the inner regions, followed by later accretion in the outskirts. Nevertheless, it is worth noting that there are substantial variations in this pattern among the Phoenix haloes. To shed light on this variability, we reference the formation redshift, denoted as $z_{\rm form}$, for these seven haloes, as obtained from \cite{Gao:2012a}. The formation redshift values are as follows: $z_{\rm form} = 1.17$ for Ph-A, $0.76$ for Ph-C, $0.46$ for Ph-D, $0.91$ for Ph-E, $1.1$ for Ph-F, $0.18$ for Ph-G, and $0.56$ for Ph-I.

For haloes Ph-D and Ph-G, the majority of particles are accreted after the redshift of $z=0.65$. In contrast, for the remaining haloes, particles in this redshift range also contribute significantly to the final halo mass, although their contributions fall short of the 50\% mark. Notably, haloes Ph-A, Ph-D, Ph-F, and Ph-G exhibit a particularly pronounced inside-out formation pattern, especially after the redshift of $z=3$, with later accreted particles predominantly located in the halo outskirts.

For haloes Ph-C, Ph-E, and Ph-I, this inside-out pattern commences at a later stage. In these cases, particles accreted at $z>3$ do not occupy the centralmost region at the present day. In the case of halo Ph-E, particles accreted at $z>3$ are offset from the halo centre. In the case of halo Ph-C, particles accreted during the redshift ranges $1.07<z<3$ and $0.65<z<1.07$ collectively contribute to nearly 50\% of the mass in the central region, while particles accreted at $z>3$ play a marginal role in shaping the halo's mass and structure.

Note that the radial profile of particles with $z_{\rm acc} > 3$ might sensitively depend on the definition of the halo centre. To check this, we have re-done Figure~\ref{fig:acc_time_profile_ph2} by setting the $z = 0$ halo centre as the centre of mass position of the particles within the half-mass radius. We found that the particles with $z_{\rm acc} > 3$ in the Ph-E halo occupy the centralmost region. However, the Ph-C and Ph-I haloes still exhibit similar profiles as in Figure~\ref{fig:acc_time_profile_ph2}. Comparing with the Aquarius haloes shown in \citetalias{Wang:2011a}, which always have their $z_{\rm acc} > 6$ particles in the centralmost region, this reflects that cluster haloes are less relaxed, and therefore it is more difficult to identify a robust centre. It might also be worth investigating the effect of using alternative halo finder algorithms, but this is beyond the scope of this study.

\begin{figure*}
    \includegraphics[width=0.9\textwidth]{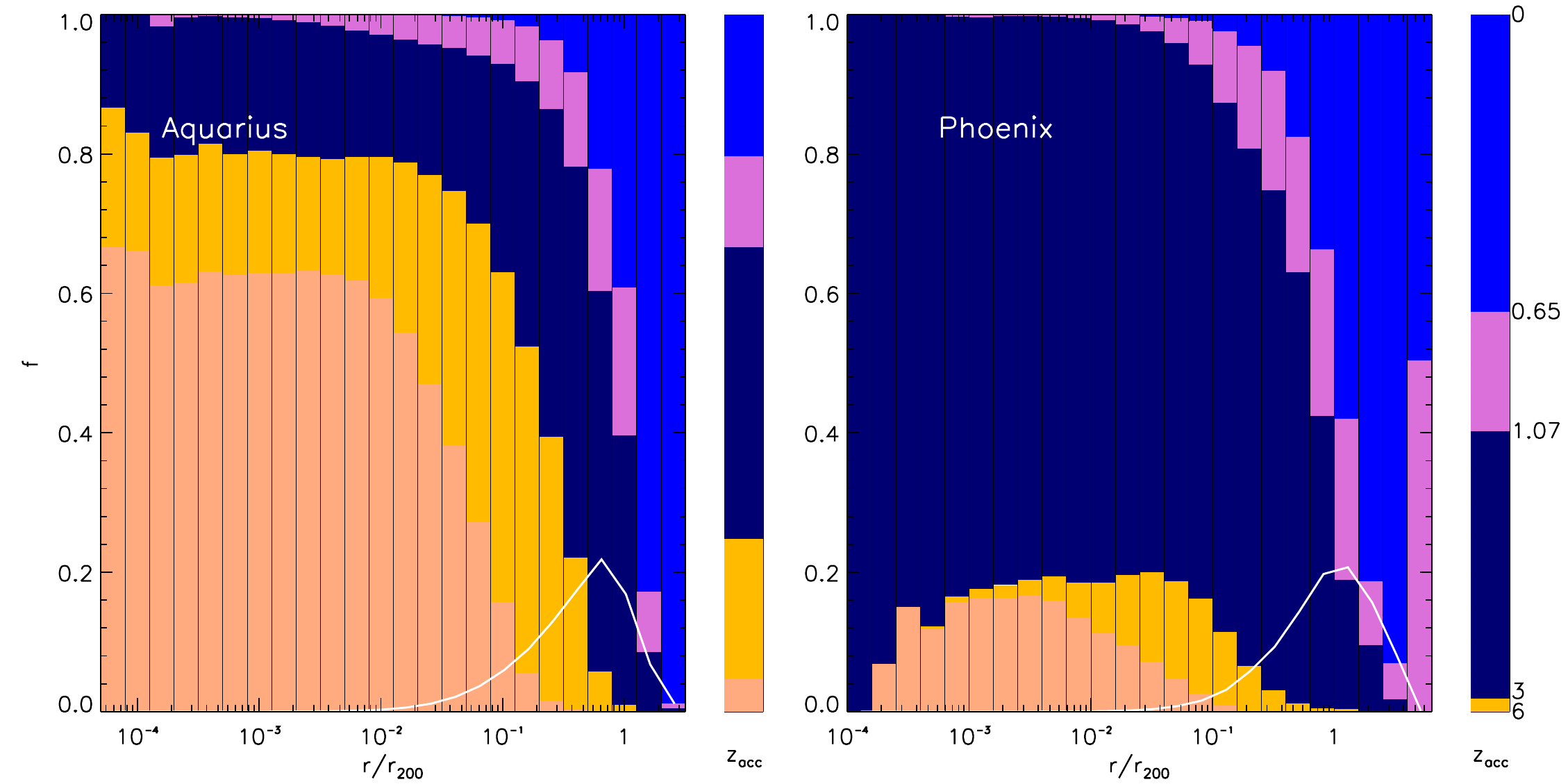}
    \caption{
	The stacking profiles of particle accretion redshifts for Aquarius (left) and Phoenix (right) haloes. The horizontal axis is the radius scaled by $r_{200}$, while the vertical axis is the fraction of particles accreted in certain redshift ranges. The colour bars on the right show the colour coding for each redshift bin. The white curve in each subplot is the mass fraction in each shell to the total mass plotted. For better clarity in the colour bar, particles of $z>6$ (in beige) are not labeled.}
    \label{fig:acc_time_profile_Aq2Ph2}
\end{figure*}

In Figure \ref{fig:acc_time_profile_Aq2Ph2}, we present stacked radial particle accretion time profiles for both Aquarius and Phoenix haloes, with the radius scaled by the virial radius of each respective halo. In a general sense, both Aquarius and Phoenix haloes exhibit qualitatively similar inside-out accretion patterns. For Aquarius haloes, a stable central region has already formed before the redshift of $z=6$ and has persisted to the present day. Conversely, for Phoenix haloes, a significant portion of the mass in the central regions is accreted after the redshift of $z=3$. This difference highlights the varying temporal evolution of the central regions in these two sets of haloes.

\begin{figure}
    \centering
    \includegraphics[width=\columnwidth]{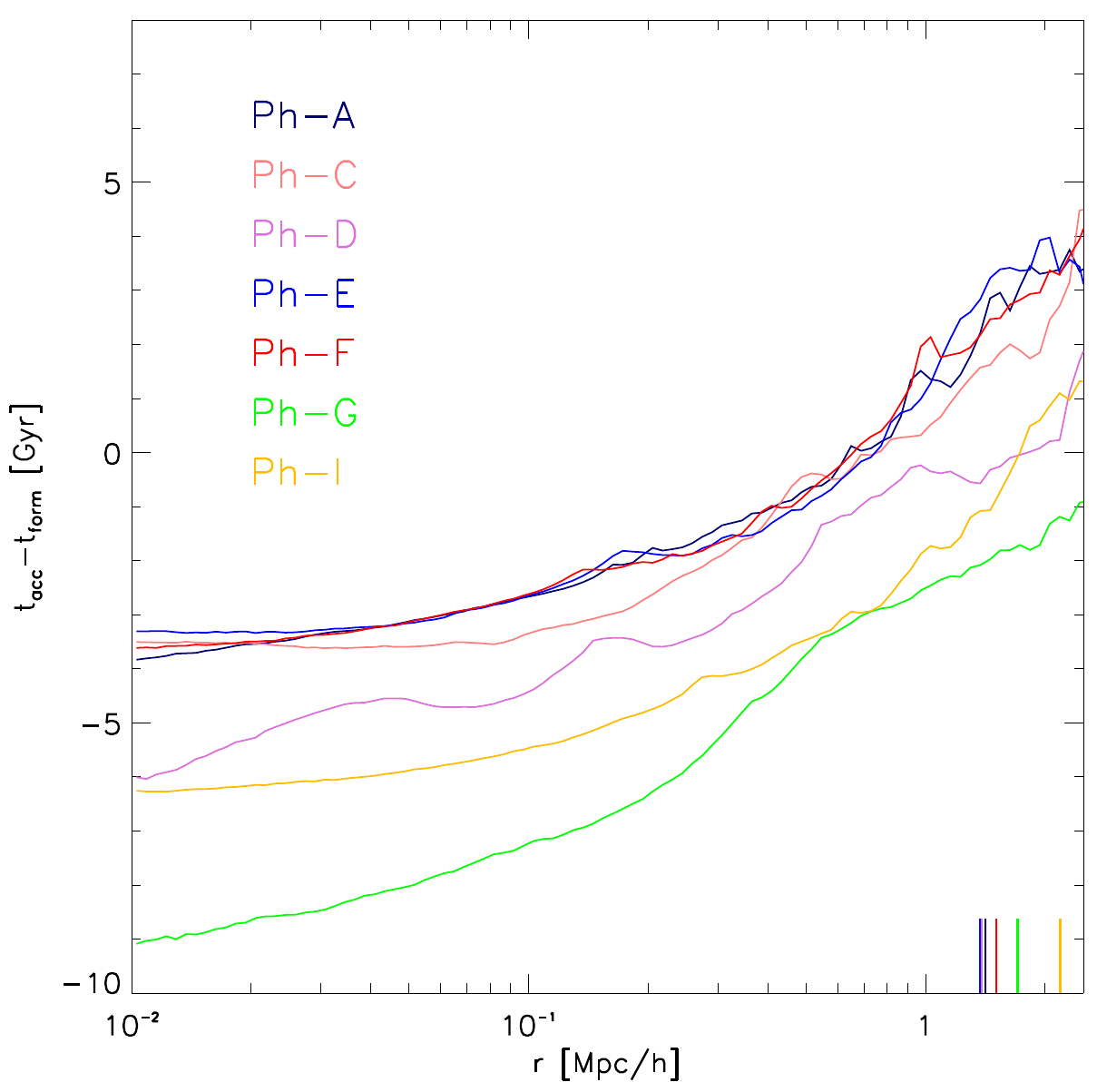}
    \caption{
	Radial profiles of mean particle accretion times at different radii for Phoenix haloes. Vertical segments indicate their virial radii. Note that the accretion time is normalized by the formation time of the halo.
	}
    \label{fig:accretion_time_ph2_norm_cases}
\end{figure}

Given that cluster haloes tend to have formation redshifts later than galactic haloes, we introduce the concept of the normalized accretion time, denoted as $t_{\rm acc} - t_{\rm form}$. Here, $t_{\rm acc}$ represents the cosmic time, or equivalently the age of the universe, corresponding to $z_{\rm acc}$, while $t_{\rm form}$ corresponds to $z_{\rm form}$. In Figure \ref{fig:accretion_time_ph2_norm_cases}, we present the radial profiles of mean normalized particle accretion times for all seven Phoenix haloes. These profiles can be effectively categorized into two distinct groups. Haloes Ph-D, Ph-G, and Ph-I exhibit behaviours distinct from the other four haloes, which exhibit more similar patterns. This observation aligns with expectations, as the formation times of the former three haloes are later than those of the latter four. Furthermore, when comparing the normalized accretion time profiles of Phoenix haloes to those of Aquarius haloes (as shown in Figure~5 of \citetalias{Wang:2011a}), it becomes evident that Phoenix haloes tend to exhibit larger halo-to-halo variations in their accretion time profiles.

\begin{figure}
\centering
    \includegraphics[width=0.9\columnwidth]{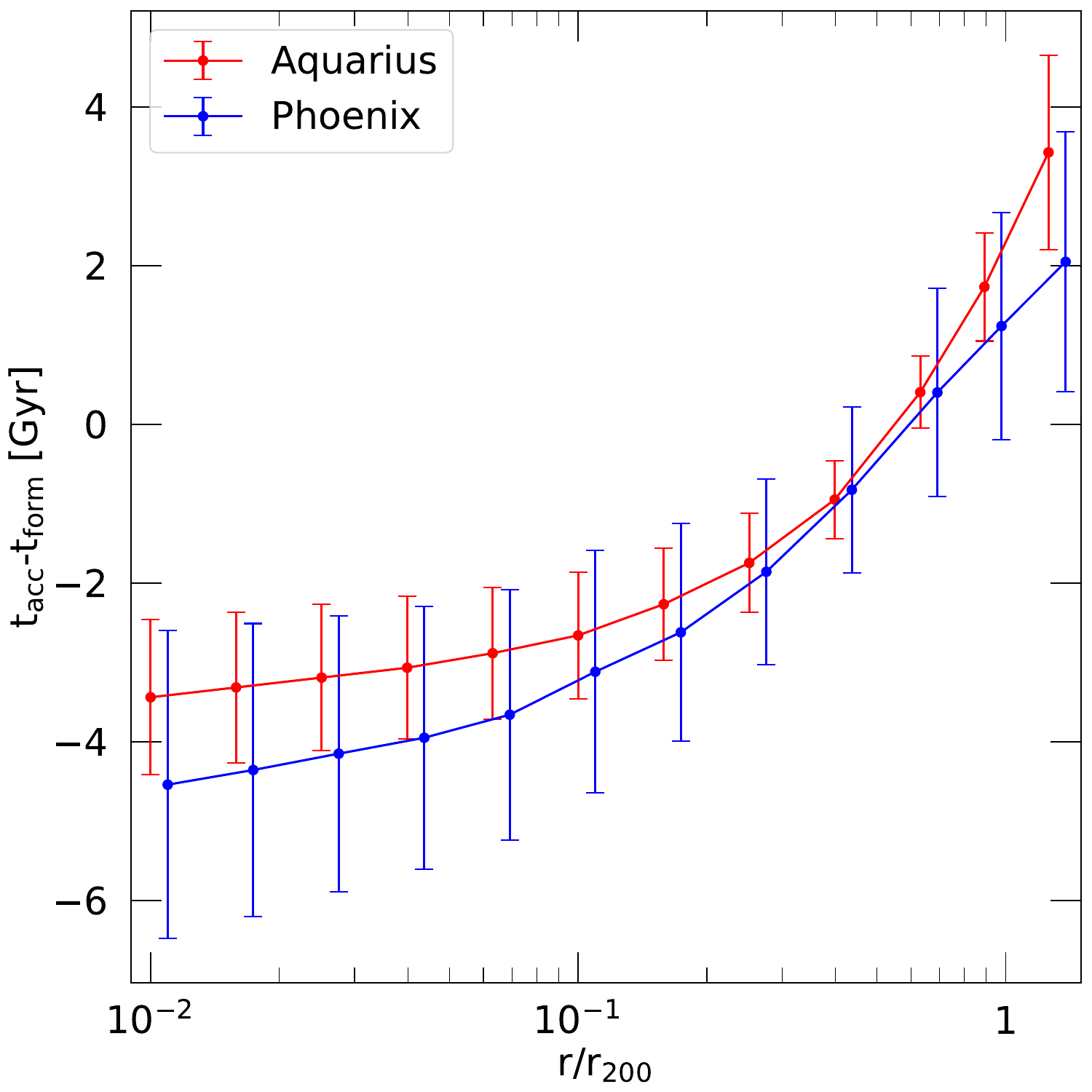}
    \caption{
    The averaged radial profile of particle accretion times for both Aquarius (red) and Phoenix (blue) haloes, normalized by the halo formation time. Error bars show the standard deviations among haloes. The line for Phoenix is right shifted slightly for better illustration.}
    \label{fig:t_acc_form_Aq2Ph2}
\end{figure}

In Figure \ref{fig:t_acc_form_Aq2Ph2}, we present a comparison of stacked $t_{\rm acc} - t_{\rm form}$ profiles averaged across six Aquarius haloes and seven Phoenix haloes. This comparison reveals that, in contrast to galactic haloes, cluster haloes tend to accumulate their innermost particles at an earlier stage relative to the halo formation time $t_{\rm form}$. Conversely, they tend to accrete their outermost particles within a shorter time interval in relation to $t_{\rm form}$.

\section{Density profiles and inner structures}\label{sec:structure}

In this section, we examine of the structural characteristics of Phoenix haloes by analyzing the evolution of density profiles and inner enclosed mass profiles. To streamline our analysis for efficiency and convenience, we opt to utilize the level-4 simulations. For readers interested in numerical convergence studies pertaining to Phoenix haloes, we direct their attention to the work conducted by \citet{Gao:2012a}.

\subsection{Evolution of profile}

\begin{figure*}
    \includegraphics[width=0.9\textwidth]{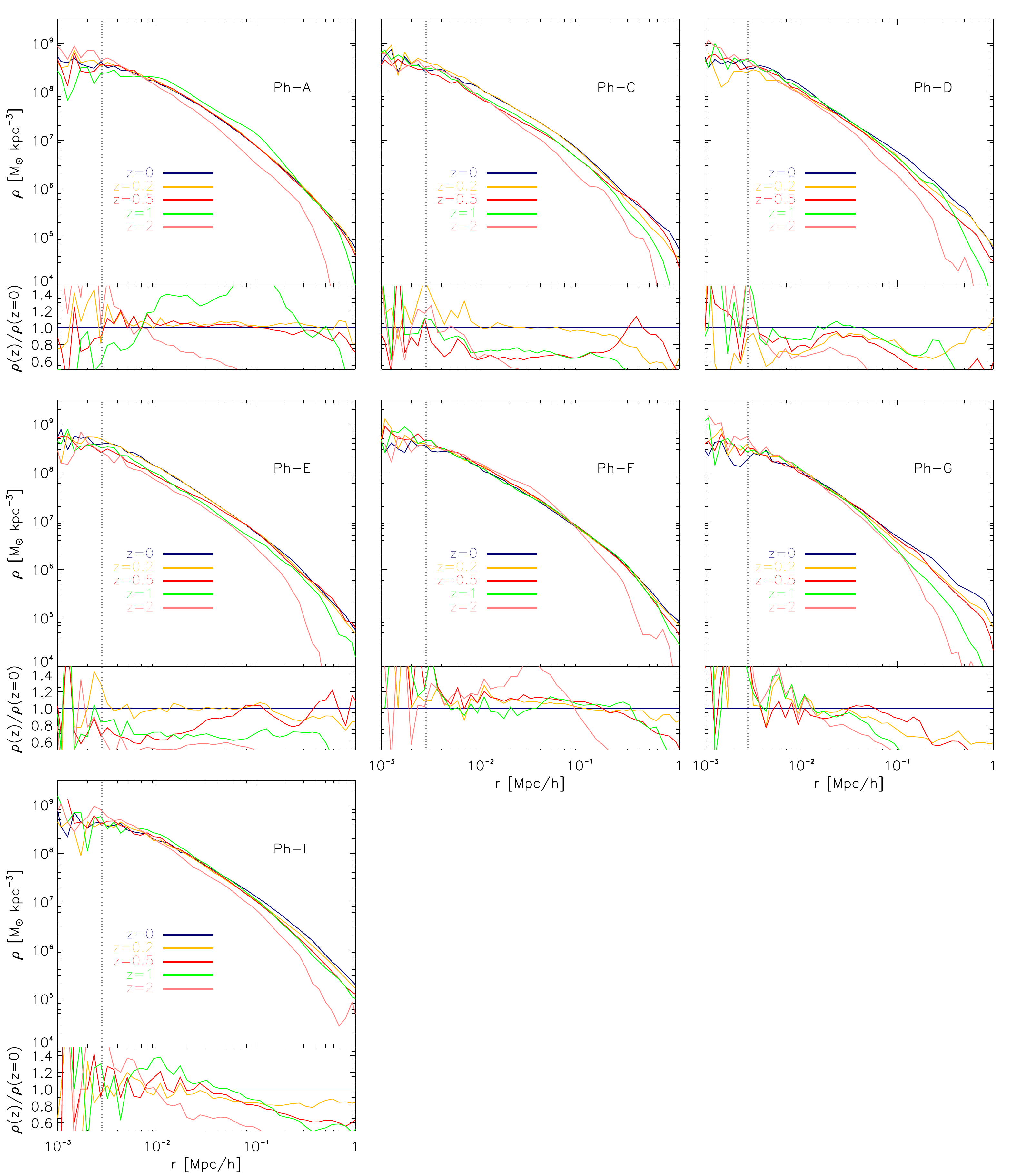}
    \caption{
    Halo density profiles at redshifts of 0, 0.2, 0.5, 1, 2. The bottom panels show the ratio of each density profile to that of the redshift 0. The dotted vertical lines give the softening length of the Phoenix haloes, indicating the reliable radial range. Generally speaking, profiles are not stable even at very low redshifts. 
    }
    \label{fig:density_prof}
\end{figure*}

We closely examine the evolution of density profiles within Phoenix haloes spanning from redshift 2 to 0. In Figure \ref{fig:density_prof}, we present the density profiles within a radius of $1 \ h^{-1}{\rm Mpc}$ for all seven Phoenix haloes at redshifts 2, 1, 0.5, 0.2, and 0. The upper panels feature the physical halo radius on the horizontal axis and the spherically-averaged density in units of $\rm M_\odot\,kpc^{-3}$ on the vertical axis. The lower panels depict the ratio between the density profiles at different redshifts and the density profile at $z = 0$.

Notably, the density profiles exhibit remarkably strong evolution, and substantial halo-to-halo variations. As previously mentioned, Ph-A and Ph-F have formation times ($z_{\rm form}>1$) earlier than the other Phoenix haloes. Correspondingly, the density profiles for Ph-A and Ph-F display greater stability than the other five haloes. In the case of Ph-A, the density profile undergoes minimal evolution after reaching a redshift of $z = 0.5$. The profile ratios at $z = 0.5$ and $0.2$ hover around 1, signifying a stable structure. Similarly, the density profile of Ph-F undergoes very little evolution after reaching a redshift of $z = 1$, particularly at radii $r < 500 \ h^{-1}{\rm kpc}$. In contrast, all other haloes demonstrate dramatic evolution in their density profiles. In fact, some of them (Ph-D, Ph-G, and Ph-I) do not exhibit stable density profiles even at a redshift as low as $0.2$. The pronounced halo-to-halo variations are consistent with the significant scatter in their formation redshifts.

\subsection{Inner structures}

\begin{figure*}
    \includegraphics[width=0.9\textwidth]{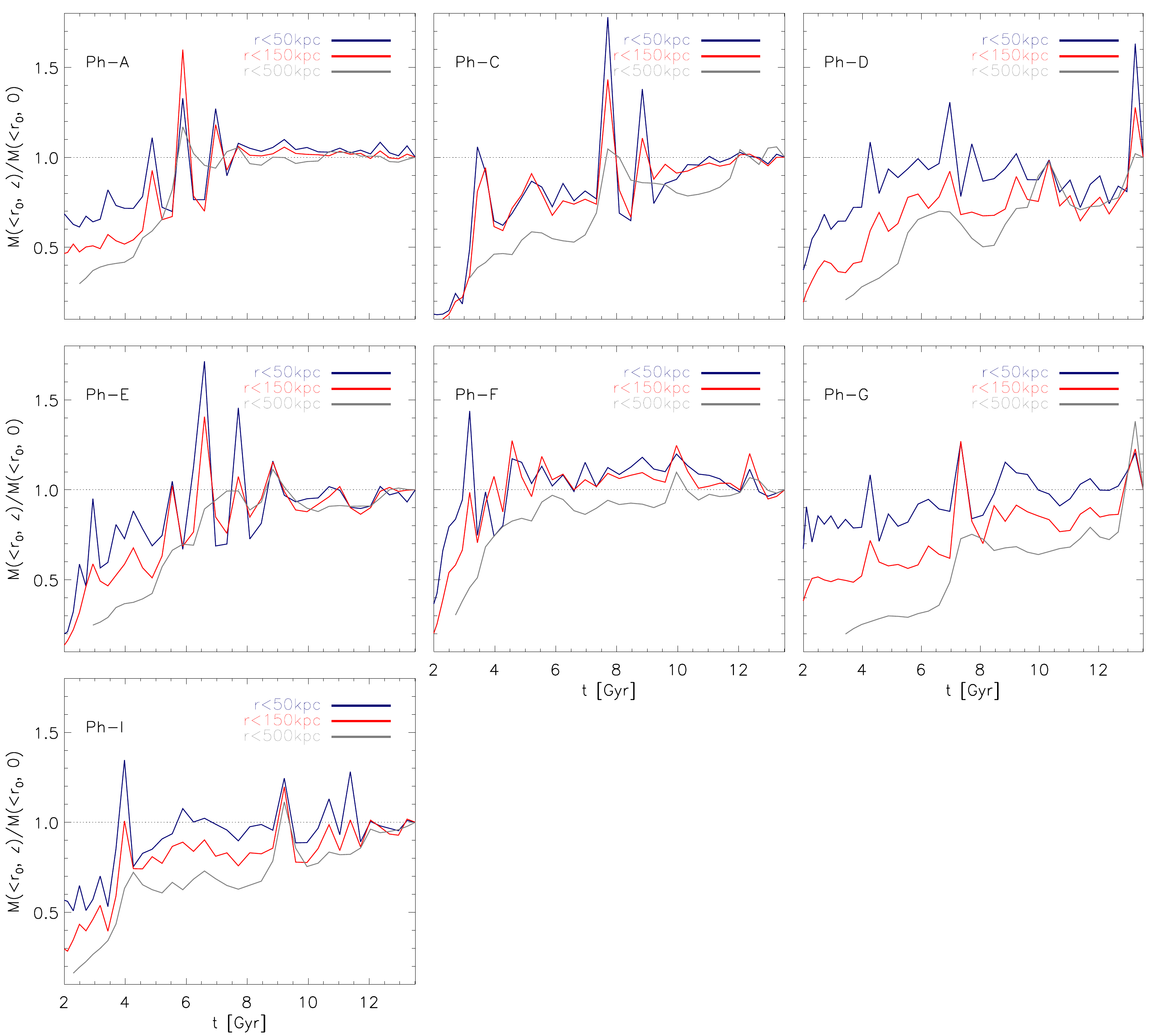}
    \caption{
	The mass enclosed within 50, 150 and 500~kpc as a function of time for the Phoenix haloes. The enclosed mass has been normalized to that at redshift 0. Masses within the inner-most 50~kpc and 150~kpc are considered, and mass within 500~kpc is overplotted for reference. The breaks in the 500~kpc curves are due to the fact that we do not count snapshots if $r_{200} < 2R_0$. For almost all the haloes, there are dramatic spikes and troughs. This is a sign that Phoenix haloes suffer from active mass changes. 
	}
    \label{fig:particle_evo}
\end{figure*}

We now turn our attention to investigating the inner particle distribution and its temporal evolution. To accomplish this, we focus on two central spheres with radii of $R_0=50 \ {\rm kpc}$ and $R_0 = 150 \ {\rm kpc}$, respectively. We trace the enclosed mass within these spheres from $z = 0$ until the virial radius of the Phoenix halo progenitor becomes smaller than $2R_0$, i.e., when $r_{200}(z) < 2 R_0$. The results of this analysis are displayed in Figure \ref{fig:particle_evo}, where the horizontal axis represents the age of the Universe, and the vertical axis depicts the ratio between the enclosed mass at various redshifts and that at $z = 0$. Different coloured lines in the legend represent the evolution of the enclosed mass within different spherical regions.

For reference, we also overlay the results for $R_0 = 500 \ {\rm kpc}$. Notably, this analysis reveals intriguingly distinct behaviours when compared to the Aquarius haloes studied in \citetalias{Wang:2011a}. Aquarius haloes generally exhibit stable inner structures after approximately 7 Gyr of the Universe's age\footnote{A similar investigation of the evolution of enclosed mass at different physical radii was conducted by \citet{Diemand:2007} using a Milky Way-sized galaxy from the Via Lactea simulation, and the authors reached a similar conclusion.}, with the exception of Aq-F, which experiences a major merger very recently.

In contrast, all Phoenix haloes display substantial evolution in their inner mass distribution after 7 Gyr, and some haloes, such as Ph-D and Ph-G, even undergo fairly active changes in the recent past. Once again, the significant halo-to-halo variation is evident.

\section{Discussion}\label{sec:discussion}

\begin{figure*}
    \includegraphics[width=\textwidth]{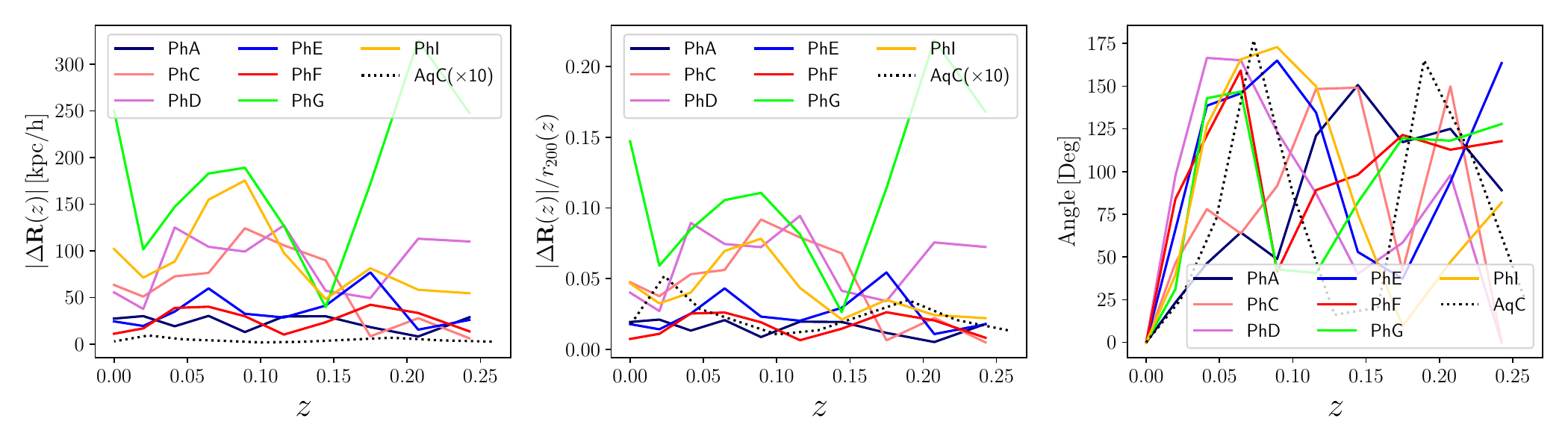}
    \caption{The vector constructed by the mostbound centre and half mass centre, at different redshift. Left: the norm of the vector. Middle: the norm of the vector scaled by the $r_{200}$ at the redshift. Right: the vector angles relative to that of redshift 0. For better illustration, the norm values in the left and middle plots for Aq-C have been scaled up by 10.}
    \label{fig:vector_z}
\end{figure*}

The more active and younger age of cluster haloes, especially the inner regions, could explain some observational facts of cluster haloes. On the estimation of mass of cluster haloes, normally one needs the anchored potential centre on the brightest cluster galaxy (BCG). For cluster simulations this is usually not a problem, as the cuspy potential centre is able to fix the BCG in the centre. However, there is an ongoing debate of whether this central galaxy paradigm (i.e. BCG locates in the centre of cluster) is valid. Recent observational analysis, such as in \citet{Lopes:2018a} and HST data used in \citet{Harvey:2017a}, have shown offsets of BCG to the cluster centre. For \citet{Harvey:2017a} especially, they use a simple harmonic oscillator to estimate the dynamical wobbling of BCG within the centre, and finding a non-zero magnitude of $A_w = 11.82$ kpc. In \citet{Kim2:2017a} this offset can be explained as a self-interact dark matter model. But in this work, we suggest that the overall and inner region's degree of relaxation can be considerably different. One must assume the dynamic state is well virialized when trying to estimate the density profile with a dynamical method. However, as we demonstrated in the previous section, the $z = 0$ central regions can actively evolve and might not be well-relaxed.

The lack of well-defined virialization has a notable impact on the determination of the halo centre. This effect is demonstrated through the following experiment, and the illustration is presented in Figure~\ref{fig:vector_z}. In this experiment, for each halo, we designate their potential centre at the certain redshift, which is the default halo `centre' used in the simulations, as $\bmath{R}_1(z)$. We then calculate the mass centre of the particles within the half-mass radius, denoted as $\bmath{R}_2(z)$. Here, $\bmath{R}_2(z)$ is computed using an iterative approach, i.e. we start with the potential centre and then iteratively update the centre as the mass centre of the particles within the half-mass radius until the centre eventually converges. The goal is to evaluate the difference between these two centres. To achieve this, we calculate the vector difference between the two, represented as $\Delta \bmath{R}(z) = \bmath{R}_2 (z) - \bmath{R}_1 (z)$, for the last 10 snapshots. The magnitude (or norm) of the vector, $|\Delta \bmath{R}(z)|$, and the $r_{200}(z)$ scaled version, $|\Delta \bmath{R}(z)| / r_{200} (z)$, are depicted in the left and middle plots. To provide a clearer visualization, we anchor the vector direction at the $z=0$ snapshot and calculate the angles between this vector and the other vectors at $z > 0$, as shown in the right figure. For comparison, we include the same analysis for Aquarius halo C as well. For better illustration, Aq-C's values for vertical axis are multiplied by 10 in the left and middle plots. It is worth noting that, in general, the disparities between the two definitions are as large as $\sim 50\,h^{-1}{\rm kpc}$. They easily exceed 2\% $r_{200}$ and may even reach approximately 10\% $r_{200}$. However, for the Aquarius halo they are at $\sim 0.3\,h^{-1}{\rm kpc}$ level and only $\sim 0.1\%$ of $r_{200}$. The right plot illustrates that the difference is not a simple shift between the two vectors but involves a more intricate `move-around' pattern. Note that in this plot, an angle of 90$^{\circ}$ defines the plane perpendicular to the initial vector.

In order to have a clear or even full understanding of these issues, baryon effects should also be considered, and the high-resolution hydrodynamic simulation of cluster haloes would be suitable. We note that in recent years there have been many progress in galaxy cluster simulations including baryons \citep[e.g.][]{Planelles:2013a,Sembolini:2013a,Rasia:2015a,Wu:2015a,Bahe:2017a,Barnes:2017b,Barnes:2017a,Cui:2018a,Henden:2018a,Nelson:2024}. A combination of the methodology used in our work with these hydrodynamic simulations should bring additional insights on the mass accretion history of clusters during cosmic time.

\section{Summary}\label{sec:summary}

In this work, we study the accretion histories and structures of cluster haloes using the Phoenix zoom-in dark matter-only simulations, and compare with galactic haloes from the Aquarius simulations. We summarize our work and findings as follows:

\begin{itemize}

\item For the mass growth history, Phoenix haloes grow more actively at $z<1$. The formation time of cluster-sized Phoenix haloes is in general later than that of Milky Way-sized Aquarius haloes, which is consistent with many previous studies.
\item Following the analysis in \citetalias{Wang:2011a}, we trace the history of each particle in each $z = 0$ Phoenix halo and register the redshift when this particle was first accreted onto the main progenitor of the Phoenix halo, as well as the mass of the halo which brought this particle into the Phoenix progenitor. We then study the radial profile of the particle merger mass ratios and accretion redshifts.
 \begin{enumerate}
 \item In general, the Phoenix haloes exhibit very similar profiles as the Aquarius haloes, i.e. particles that were accreted early and those that were brought from relatively major mergers dominate the mass in the central region, while the particles that were accreted via diffuse accretion and minor mergers dominate the mass in the outskirts. Compared to galactic haloes, cluster haloes tend have less contribution from $z_{\rm acc}>6$ particles in the inner regions, implying that their inner regions are more affected by mass accretion at lower redshifts. Particles accreted via major (minor) mergers make larger contributions in the central (outskirt) regions of cluster haloes compared to those in galactic haloes.
 \item Both cluster and galactic haloes experience inside-out `onion-like' growth, with mass in the inner regions being accreted early whereas the mass in the outskirts is accreted later.
 \end{enumerate}
\item When looking at density profiles and enclosed masses of inner regions at different redshifts, we see that there is a much stronger evolution for cluster haloes compared to galactic haloes. This implies that cluster haloes are less stable than galactic haloes, especially in the inner regions.
\end{itemize}

\section*{Acknowledgements}

We thank the anonymous referee for a constructive and useful report that helped to significantly improve our manuscript. This work is supported by the National Key R\&D Program of China (2022YFA1602901  and Nos. 2018YFE0202900), the NSFC grant (Nos. 11988101, 11873051, 12125302, 11903043), CAS Project for Young Scientists in Basic Research Grant (No. YSBR-062), and the K.C. Wong Education Foundation.

\section*{Data availability}
The simulation data used in this article will be shared upon a reasonable request to the corresponding author.

\bibliographystyle{mnras}
\bibliography{ref}


\bsp	
\label{lastpage}
\end{document}